\documentstyle[12pt]{article}

\def \invisible{\mbox{$\rule{0mm}{1mm}$}}
\def \mathbox(#1){\invisible\ifmmode{{#1}}\else{\mbox{${#1}$}}\fi}
\def \mbf(#1){\mbox{\boldmath{$#1$}}}
\def\mbfsig{\mbf(\sigma)}
\def\mbfp{\mbf(p)}

\def\calO{{\cal O}}
\def\calA{{\cal A}}
\def\calS{{\cal S}}
\def\Del{\Delta}

\def\MA{{\tiny  {\rm MA}}}
\def\MS{{\rm {\small MS}}}
\def\rttwo{\sqrt{2}}

\def\lamlam{\lam_i\cdot\lam_j}
\def\lam{\lambda}

\def\psiL {\psi_L}
\def\psibR{\psib_R}
\def\psib{\bar\psi}

\def\piii{p_{\rm III}}
\def\qbar{\bar q}

\def\mbfq{\mbf(q)}

\def\sigsig{\sigv_i\cdot\sigv_j}

\def\sig{\sigma}

\def\bra{\langle}
\def\ket{\rangle}

\def\III{\mbox{\rm I\thinspace I\thinspace I}}

\def\HIII{\mathbox(H_{\rm III})}
\def\VIII{\mathbox(V_{\rm III})}

\def\Voge{\mathbox({V_{\rm OGE}})}

\def\forth(#1){\mathbox(\frac{#1}{4})}
\def\lamlam{\lam_i\cdot\lam_j}

\def\sigsig{\mbfsig_i\cdot\mbfsig_j}

\def\FRAC#1#2{\leavevmode\kern-.em
\raise.5ex\hbox{\the\scriptfont0 #1}\kern-.em
/\kern-.15em\lower.25ex\hbox{\the\scriptfont0 #2}}

\def\barA{{\overline A}}
\def\barDel{{\overline \Delta}}
\def\barchi{\;{\overline \chi}}
\def\bartheta{\;{\overline \theta}}
\def\barcalO{{\overline {\cal O}}}
\def\alphas{\alpha_s}
\def\mbfr{\mbf(r)}
\def\Vzerotwo{V_{0}^{(2)}}
\def\leftw{\left\{\rule{0pt}{0.5cm}\right.}
\def\rightw{\left.\rule{0pt}{0.5cm}\right\}}
\newif\ifrefphysrev

\refphysrevtrue

\def \vol(#1,#2,#3){\ifrefphysrev{{\bf {#1}},
{#3} (19{#2})}\else{{{\bf {#1}}(19{#2}){#3}}}\fi}
\def \NP(#1,#2,#3){Nucl.\ Phys.\          \vol(#1,#2,#3)}
\def \PL(#1,#2,#3){Phys.\ Lett.\          \vol(#1,#2,#3)}
\def \PRL(#1,#2,#3){Phys.\ Rev.\ Lett.\   \vol(#1,#2,#3)}
\def \PRp(#1,#2,#3){Phys.\ Rep.\          \vol(#1,#2,#3)}
\def \PR(#1,#2,#3){Phys.\ Rev.\           \vol(#1,#2,#3)}
\def \PTP(#1,#2,#3){Prog.\ Theor.\ Phys.\ \vol(#1,#2,#3)}
\def \ibid(#1,#2,#3){{\it ibid.}\         \vol(#1,#2,#3)}

\makeatletter
\def\scriptsize{\@setsize\scriptsize{14.5pt}\xipt\@xipt
\abovedisplayskip 11\p@ plus3\p@ minus6\p@
\belowdisplayskip \abovedisplayskip
\abovedisplayshortskip  \z@ plus3\p@
\belowdisplayshortskip  6.5\p@ plus3.5\p@ minus3\p@
\def\@listi{\leftmargin\leftmargini
\parsep 4.5\p@ plus2\p@ minus\p@ \itemsep \parsep
\topsep 9\p@ plus3\p@ minus5\p@}}
\def \@magscale#1{ scaled \magstep #1}
\font\frtnsfb = cmssbx10 \@magscale2 
\def \half(#1){\mathbox(\frac{#1}{2})}
\def \ninej(#1,#2,#3,#4,#5,#6,#7,#8,#9){\mathbox(\left\{\matrix
     {#1&#2&#3\cr#4&#5&#6\cr#7&#8&#9\cr}\right\})}
\newif\ifnoncomplete

\def\final{\noncompletefalse\typeout{** FINAL form}}
\noncompletefalse

\def\@cite#1#2{\unskip\nobreak\relax
    {[#1]}} 

\def\citenum#1{{\def\@cite##1##2{##1}\cite{#1}}}
\def\citea#1{\@cite{#1}{}}

\newcount\@tempcntc
\def\@citex[#1]#2{\if@filesw\immediate\write\@auxout{%
\string\citation{#2}}\fi
  \@tempcnta\z@\@tempcntb\m@ne\def\@citea{}\@cite{\@for\@citeb:=#2\do
    {\@ifundefined
       {b@\@citeb}{\@citeo\@tempcntb\m@ne\@citea\def\@citea{,}%
{\bf ?}\@warning
       {Citation `\@citeb' on page \thepage \space undefined}}%
{\setbox\z@\hbox{\global\@tempcntc0\csname b@\@citeb\endcsname\relax}%
     \ifnum\@tempcntc=\z@ \@citeo\@tempcntb\m@ne
       \@citea\def\@citea{,}\hbox{\csname b@\@citeb\endcsname}%
     \else
      \advance\@tempcntb\@ne
      \ifnum\@tempcntb=\@tempcntc
      \else\advance\@tempcntb\m@ne\@citeo
      \@tempcnta\@tempcntc\@tempcntb\@tempcntc\fi\fi}}\@citeo}{#1}}
\def\@citeo{\ifnum\@tempcnta>\@tempcntb\else\@citea\def\@citea{,}%
  \ifnum\@tempcnta=\@tempcntb\the\@tempcnta\else
   {\advance\@tempcnta\@ne\ifnum\@tempcnta=\@tempcntb %
\else \def\@citea{--}\fi
    \advance\@tempcnta\m@ne\the\@tempcnta\@citea\the\@tempcntb}\fi\fi}

\def\affiliation#1{\gdef\@affiliation{#1}}

\def\and{\cr \makebox[0in]{\rule[-1cm]{0mm}{1cm}and } \cr}

\def\maketitle{\par
 \begingroup
 \def\thefootnote{\fnsymbol{footnote}}
 \def\@makefnmark{\hbox
 to 0pt{$^{\@thefnmark}$\hss}}
 \if@twocolumn
 \twocolumn[\@maketitle]
 \else \newpage
 \global\@topnum\z@ \@maketitle \fi\thispagestyle{plain}\@thanks
 \endgroup
 \setcounter{footnote}{0}
 \let\maketitle\relax
 \let\@maketitle\relax
 \gdef\@thanks{}\gdef\@author{}\gdef\@title{}
 \gdef\@affiliation{} \let\affiliation\relax	%
 \let\thanks\relax}

\def\@maketitle{\newpage
 \null
 \vskip 0em plus 2em minus 0em     
 \ifx\@date\@empty\else
   \begin{flushright}
    {\ifnoncomplete(\today)
     \else{{\normalsize \@date}\\}\fi}      
   \end{flushright}
   \vskip 3em plus 2em minus 2em   
 \fi
 \begin{center}
  {\frtnsfb \@title \par}     
  \vskip 3em plus 1em minus 1.5em  
  {
   \lineskip .5em plus 0em minus .3em   
   \begin{tabular}[t]{c}\@author\\
   \end{tabular}\par}
  \vskip 0.5em plus 1em minus 1.5em  
  { \sl \@affiliation \par}
\end{center}
 \par
 \vskip 6em plus 2em minus 4em}     

\def\abstract{\if@twocolumn
\section*{Abstract}
\else \normalsize
\fi}

\def\endabstract{\if@twocolumn\fi\par\clearpage}

 \makeatother
\final
\refphysrevtrue

\begin{document}
\title{Effects of Instantons on the Excited Baryons \\
and Two-Nucleon Systems}
\author{Sachiko Takeuchi}
\affiliation{
Department of Public Health and Environmental Science,\\
School of Medicine,
Tokyo Medical and Dental University,\\
1-5-45  Yushima, Bunkyo, Tokyo 113, Japan}
\date{\today}
\maketitle
{\bf Abstract:~~}
The effects of the spin-orbit and the tensor parts of
the instanton-induced interaction
on the excited nonstrange baryons up to $N$=2
and on the two-nucleon systems are investigated.
The spin-orbit force from the instanton-induced interaction
cancels most of that from the one-gluon exchange
in the excited baryons
while the spin-orbit force in the nucleon systems remains strong
after the inclusion of the instanton-induced interaction.
The model including the spin-orbit and the tensor terms
of the instanton-induced interaction
as well as the one-gluon exchange is found to reproduce successfully
the excited baryon
mass spectrum and the scattering phase shifts of two nucleons
in the spin-triplet
relative $P$-wave state.


\newpage

\section{Introduction}

Valence quark models
have been applied to low-energy light-quark systems and found to be
successful in
reproducing major properties of the hadrons and hadronic systems.
The reason why
such models can be so successful
is not well understood.
The empirical approach, however, suggests a few reasons.
One of them is that the model space has an appropriate symmetry.
Another reason is that a light quark has rather heavy
effective mass.
In such a low energy region,
the effects of
complicated configurations such as $q\qbar$ excitation and
 dynamical gluon effects are
considered to be taken into account by employing
constituent valence quarks and effective interactions among quarks
with the required symmetry.

These empirical models usually
contains three terms:
the kinetic term, the confinement term, and the
effective one-gluon exchange (OGE) term.
It is considered that OGE stands for the
perturbative gluon effects and that the confinement force represents
the long-range nonperturbative gluon effects.

It is well known
that the color magnetic interaction (CMI) in OGE
is responsible to produce many of the hadron properties.
By adjusting the strength of OGE,
CMI can reproduce
the hyperfine splittings (HFS,
e.g., ground state N-$\Delta$ mass difference)
 \cite{GS76,IK78,Is92,SS90}
as well as the short range repulsion of
the two-nucleon systems in the relative $S$-wave
\cite{MYO84,TSY89,Sh89}.
It, however, is also known that
the strength of OGE determined in this
empirical way is much greater than 1, which makes it hard
to treat it as the perturbative effect.

Moreover, the valence quark model including only OGE
as an origin of HFS
has a spin-orbit problem.
The spin-orbit part of OGE is strong; it is just strong enough
to explain the observed large spin-orbit force between two nucleons
\cite{MYO84,TSY89}.
On the other hand,
the experimental mass spectrum of the excited baryons,
N$^*$ and $\Delta^*$
resonances, indicates
that such a strong spin-orbit force should not exist between quarks.
A valence quark model in which the spin-orbit parts
of the quark-quark
 interaction are removed by hands
can well simulate the  observed mass spectrum
\cite{GS76,IK78,Is92,SS90}.
It was pointed out that the confinement force also
produces the spin-orbit force, which may cancel the one from OGE
in the excited baryons \cite{IK78}.
Suppose one takes the spin-orbit part of a
 two-body confinement force into account, however,
it also cancels the spin-orbit part of OGE
in the nuclear force \cite{MYO84}.
To explain both of the features at the same time
is highly nontrivial.

The instantons were originally introduced in  relation to
the $U_A$(1) problem;
their coupling to the surrounding light-quark zero modes
produces the flavor-singlet interaction between quarks,
which leads the observed large mass difference
of $\eta'$-$\eta$ mesons \cite{tH76,Sh84}.
How this instanton-induced interaction (\III),
which contributes by a few hundred MeV in the
meson sector, affects in other hadron systems
is an interesting problem.
Actually, several recent works
indicate  that the effect is also large
in the baryon sector\cite{Ko85,OT91,TO91,Ta94,MT91,Negele}.
It is hard to determine the strength of the instanton effects
quantitatively directly from QCD.
The above empirical works on \III\ can contribute  also in
understanding the structure of the QCD vacuum.
We argue that a valence quark model should include \III\ as a
short-range nonperturbative gluon effect
in addition to the other aforementioned gluon effects.

In ref.\ \cite{Ta94}, we demonstrated that
introducing \III\ may solve the above
difficulty in the $P$-wave systems due to  the cancellation between
OGE and \III.
In this paper, we discuss
effects of the noncentral parts, especially the spin-orbit part of
\III\ on the excited nonstrange baryons up to $N=2$
and on the two-nucleon systems.
 We employ a quark potential model  for that purpose
because it can deal both with the single baryons and
with the two-nucleon systems, and because the discussion
based on the symmetry can be performed more clearly
for the present subject.
In section 2, we will show the model hamiltonian.
The discussion based on the symmetry is presented in section 3.
The numerical results are shown in section 4.
The discussion and the summary are in section 5.
The complicated wave functions and the matrix elements
are summarized in the appendix.

\section{Quark Model with Instanton Induced Interaction}

We assume that both of OGE and \III\
are included in the quark model hamiltonian:
\begin{eqnarray}
H_{\rm quark}
&=&
 K+(1-\piii)V_{\rm OGE} + \piii \VIII\ + V_{\rm conf}\, ,
\label{eq:quarkmodel}
\end{eqnarray}
where $\piii$ is a parameter which represents the rate of
the $S$-wave N-$\Delta$
mass difference explained by \III.
 When one introduces the
interaction strong enough to give the observed $\eta$-$\eta'$ mass
difference, $\piii$ becomes 0.3--0.4
\cite{OT91,TO91,Ta94}.
$V_{\rm OGE}$ and
\VIII\
are the Galilei invariant terms of the \III\ and OGE potentials.

According to
the instanton liquid model, the size of instantons is about 0.3 fm,
which is a new scale of the low energy QCD phenomena \cite{Sh84}.
The instantons and the anti-instantons couple
to flavor-singlet light quarks.
Assuming the instanton is small enough
comparing to the system we consider,
one obtains the effective interaction between quarks
arising from that coupling
 as \cite{tH76,Ko85,OT91,TO91,Ta94}:
\begin{equation}
 \HIII = V_0^{(2)}  \sum_{i<j}
   \psibR(i) \psibR(j) \,
             {15 \over 8} {\cal A}^{flavor}_{ij}
                ( 1- {1 \over 5} \sigsig )
                          \,\psiL(j) \psiL(i)
                          + \hbox{(h.c.)} , \hfill
\label{relaIII}
\end{equation}
where $V_0^{(2)}$ is the strength of the two-body part of \III.
We obtain the following potential
performing the nonrelativistic reduction
to the lowest non-vanishing order in $(p/m)$
for each operator of different spin structure \cite{Ta94}:
\begin{eqnarray}
\VIII &=&
 V_0^{(2)}  \sum_{i<j}{\cal A}^{flavor}_{ij}
         \left[{15\over16}\!\left(1\!-\!{1\over 5} \sigsig \right)
-{1 \over m^{2}} \!\left( {3\over4}
LS
+{q^2\over12}(1\!-\!{3\over16}\lamlam)S_{12}\right)
\right]
\\
&=& V_0^{(2)}  \sum_{i<j}
       \!\left(1+{3\over32}\lamlam+\!{9\over 32}
       \lamlam\sigsig \right)
\nonumber\\
&&-{\cal A}^{flavor}_{ij}
{1 \over m^{2}} \!\left( {3\over4}
LS
+{q^2\over12}(1\!-\!{3\over16}\lamlam)S_{12}\right)
 .
\label{eq:Hiii}
\end{eqnarray}
The same procedure for OGE leads \cite{MYO84,DGG75}
\begin{equation}
V_{\rm OGE} = 4\pi\alpha_s \sum_{i<j}{(\lamlam)\over 4}\!
         \left[{1\over q^2}-{\sigsig\over6m^{2}} 
+{3\over 2m^2q^{2}}LS
+{1\over12m^2}S_{12}
\right]
\label{eq:Hoge}\\
\end{equation}
with
\begin{eqnarray}
LS&=&
{(\mbfsig_i+\mbfsig_j)}\cdot i[\mbfq\times{(\mbfp_i-\mbfp_j)}]/4
\\
S_{ij}& =&
3(\mbfsig_i\cdot \mbfq)(\mbfsig_j\cdot \mbfq)/q^2-(\sigsig) .
\end{eqnarray}
Here ${\cal A}^{flavor}_{ij} =(1-P^{flavor}_{ij})/2$
is the antisymmetrizer in the flavor space,
$m_{u}=m_{d}\equiv m$ is a constituent quark mass,
and $\mbfq$ is the three momentum transfer.
The values of
 $m$, the strength of each interaction, $\alpha_s$
and $V_0^{(2)}$,
with the size parameter $b$ of the quark core of the baryon,
are listed in table 1.
These values are chosen as follows \cite{MYO84,TSY89,OT91,TO91,Ta94}:
the quark mass is $\FRAC{1}{3}$ of the nucleon mass;
the size parameter
$b$ is taken to be a little smaller
than the real nucleon size reflecting that
the observed baryon size has contribution from the meson cloud;
$\alpha_s$ and $V_0^{(2)}$
are determined to give the ground state
N-$\Delta$ mass difference $\mu$
\begin{equation}
{4\over 3\sqrt{2\pi}}{\alpha_s\over m_u^2b^3}
= -{9\over 4 \sqrt{2\pi}^3}{V_0^{(2)} \over b^3}
= 293 {\rm MeV} \equiv \mu~~;
\label{eq6}
\end{equation}
$a_{\rm conf}$, the strength of the confinement potential,
can be determined by
$\delta m_{N}/\delta b = 0$;
and $\piii$ is taken to give the $\eta'$-$\eta$ mass difference.

\section{Symmetry}

One of the reasons that the nonrelativistic quark model
can successfully
predict the properties of the low energy system is
that the model has an appropriate symmetry.
We discuss here whether the observed properties in
the spin-orbit force, small  in the single baryons
and large in the two-nucleon systems, can be
explained by discussion based on the symmetry.

The interactions (\ref{eq:Hiii}) and (\ref{eq:Hoge}) consist of
 operators which conserve the flavor symmetry.
Thus,
the spin-orbit or the tensor part
of the two-body interactions for a quark pair,
which requires the quark pair to be symmetric in the spin space,
can be decomposed as follows:
\begin{eqnarray}
\calO &=& \barcalO^A + \calO^S + \barcalO^S + \calO^A,
\label{eq:projop}
\end{eqnarray}
where
\begin{eqnarray}
\barcalO^A  & \equiv &
\calO\calA^{orb} \calS^{spin} \calA^{color} \calA^{flavor} \\
\calO^S  & \equiv &
\calO \calA^{orb} \calS^{spin} \calS^{color} \calS^{flavor}
\\
\barcalO^S  & \equiv &
\calO \calS^{orb} \calS^{spin} \calA^{color} \calS^{flavor}
\\
\calO^A  & \equiv &
\calO\calS^{orb} \calS^{spin} \calS^{color} \calA^{flavor}
\end{eqnarray}
with antisymmetrizers $\calA$'s and symmetrizers $\calS$'s.
For a quark pair with the relative-odd partial wave,
the first two terms in the right hand side of
eq.\ (\ref{eq:projop}) are relevant, while the last two terms are
for relative-even partial-wave pairs.
The operator with bar is for color singlet pairs,
which is relevant to the single baryons.
Those for the flavor-singlet (-octet) quark pairs
are marked by $A$ ($S$).

The noncentral term of \III\ contains
only flavor-singlet components,
 $\barcalO^A$ and $\calO^A$;
OGE has all of the component in eq.\ (\ref{eq:projop}).
Since OGE is vector-particle exchange and \III\
is alike to scalar-particle exchange,
their noncentral term has an opposite sign.
Thus, there is a spin-orbit cancellation where both of OGE
and \III\ survives.
Namely, it occurs only for the flavor-singlet quark pairs:
the odd-partial wave pairs in color-singlet systems
and the even-partial wave pairs in color-octet systems.

To see the properties of the single baryons
and of the short range part of the
two-nucleon systems,
we evaluate the energy of systems by the gaussian wave functions
where the center-of-mass motion is eliminated:
$(0p)(0s)^2$ for the negative-parity single baryons,
$(1s)(0s)^2$, $(0p)^{2}(0s)$ and
$(0d)(0s)^2$ for the positive-parity single baryons, and
 and $(0p)(0s)^5$
for the six-quark systems.

In Table 2, the contribution of matrix element
of each operator is listed (see appendix)
for the negative-parity baryon N$^{*}$($\FRAC{5}{2}^{-}$),
for the positive-parity excited baryon N$^{*}$($\FRAC{7}{2}^{+}$),
and for the six-quark state
with the relative $P$-wave two-nucleon quantum number.
Once the parameters are taken to satisfy eq.(\ref{eq6}), the
contribution can be expressed in units of $\mu$ together with the
dimensionless parameter $mb$.
The contribution of the color symmetric operator, i.e.,
the operator without a bar,
is found to be dominant in the six-quark state.
Since there is no cancellation for that operator,
the spin-orbit reduction of the relative $P$-wave two-nucleon systems
 is small.
Within the single baryon, of course,
only the operators with a bar are relevant.
The OGE-\III\
cancellation occurs in the spin-orbit part operating
on the odd-wave quark pairs
in the single baryons.
It is the
spin-orbit force between
the odd-wave quark pairs
that should disappear in the single baryon
as we will show in the next section.
Thus, we expect that this cancellation will
lead the observed properties in
the spin-orbit force.

\section{Results and Discussions}
\subsection{Single Baryons}

We investigate the mass spectrum of the excited nonstrange baryons
by a nonrelativistic quark model with the spin-spin, the spin-orbit
and the tensor parts of OGE
and \III.
The central spin-independent part has been modified.
The central part of the original model hamiltonian
eq.\ (\ref{eq:NRHamiltonian})
does not produce the correct zeroth-order splitting
seen in the positive parity $N$=2 baryons.
It is mainly because the spin-independent contact interaction
from the one-gluon
exchange is strong and has a wrong sign.
As ref.\ \cite{GS76,IK78,Is92} pointed out,
the deviation of the spin-independent force
from harmonic will be much important and be expressed by the
following parameterization.
We use the same method in ref.\ \cite{Is92}
with the same values except for a little modified $E_{0}$:
\begin{eqnarray}
H_{quark} &=& H_c + (1-\piii) \widetilde{\Voge}
+\piii \widetilde{\VIII}
\label{eq:NRHamiltonian}\\
H_c &=& E_0 + N \Omega + \delta \;U
\\
E_0 &=& 1090 ~{\rm MeV}~+\piii \; \mu/2
\\
\Omega &=& 440~{\rm MeV}
\\
\delta  &=& 400~{\rm MeV}
\\
U&=& \left\{\matrix{-1 \cr -{1\over2} \cr -{2\over5} \cr
-{1\over5} \cr 0 }\right.
 ~~{\rm for}~~({\rm D}^{SF},L^P)=
 \left\{\matrix{(56',0^+) \cr (70,0^+) \cr (56,2^+)
 \cr (70,2^+) \cr {\rm otherwize} }\right.
{}~~,
\end{eqnarray}
where $\widetilde{\Voge}$ ($\widetilde{\VIII}$) is
the spin-spin, the spin-orbit and
the tensor part of $\Voge$ ($\VIII$).

Our results in this subsection are affected by only few
parameters: $E_0$ to give the ground state energy,
$\Omega$ to give the difference among
the ground states, the $N$=1 negative-parity baryons
and the $N$=2 positive-parity baryons,
$\delta$ to split the $N$=2 baryons, $\mu$ in
 eq.(\ref{eq6}) for the hyperfine splittings,
$(m_{u}b)^2$, and $\piii$ to give the relative strength of \III.
This
estimate by the harmonic oscillator wave function is affected
only by the above combinations of the parameters listed in table 1.
The values of the parameters here are
reasonable.  $E_0$ is close to $3m_u$,
and the strength of OGE $\alpha_s(1-\piii)$
becomes smaller when $\piii$ = 0.4.

In Fig.\ 1a) and b), the mass spectra of
the negative-parity and the positive parity baryons are shown.
The ground state mass is kept 940 MeV
for the nucleon and 1240 MeV for $\Delta$ in each parameter set.
The observed mass spectrum
is shown by stars (the weighted average of the observed values)
and boxes (possible error) \cite{Data}.
The number of the stars corresponds to
reliability of existence of the states:
the four-star state means that
its existence is certain while the one-star state
means that evidence of its existence is poor.

The next right to the experiment
is the mass spectrum given by $H_c$.
This three-parameter model gives an excellent prediction for the
excited baryons.
The third spectrum is derived from the hamiltonian which contains
$H_c$ and the tensor term of OGE ($\piii=0$):
it corresponds to the one in ref.\ \cite{IK78,Is92},
where the spin-orbit term is omitted by hand.
The introduction of the tensor term
gives little change in the spectrum.
Actually, one cannot conclude that the tensor term
is necessary only from the
mass spectrum; it was included so as
to give correct decay modes \cite{KI80}.

The fourth column contains the central,
the spin-spin and the spin-orbit term with $\piii =0$.
The spectrum, especially for the nucleons,
is destroyed completely;
 which is the reason why the spin-orbit term
 had to be removed in ref.\ \cite{IK78,Is92}.
The question is, however, whether all the spin-orbit terms
should be removed or not.
The excited $\Delta^{*}$ mass spectrum
is better than the nucleons'.  There, all the quark pairs
are  in the flavor symmetric; the spin-orbit term exists only
for the relative $0d$-wave pairs.

The hamiltonian of the fifth spectrum
is the same as the fourth one except that
we remove all the spin-orbit force between
the relative $0p$-wave quark pairs,
namely, flavor-singlet pairs.
The remaining spin-orbit term affects only
the relative $0d$-quark pairs.
The excited nucleon spectrum changes drastically;
most of the spin-orbit effects,
which destroy the spectrum,
are found to come from the relative $0p$-wave quark pairs.
The remaining effects of the spin-orbit force are
still somewhat stronger than the best fit, but
the spectrum becomes much more closer to the realistic one.

As we showed in the previous section, the spin-orbit force between
the relative odd partial wave pairs reduces by introducing \III.
The sixth column corresponds to the hamiltonian
(\ref{eq:NRHamiltonian}) with
$\piii$ = 0.4;
it includes the central (parameterized),
the spin-spin, the spin-orbit
and the tensor terms of both OGE and \III.
In this choice, the strength of the spin-orbit force between
$0p$-pairs reduces by 0.32 from the $\piii =0$ case
while that between $0d$-pairs reduces by $(1-\piii)=0.6$.
Note that this simple model does
only have six parameters ($E_0$, $D_0$, $\delta$, $\mu$, $(m_ub)^2$
and $\piii$) for the whole nonstrange baryons up to the
$N$=2 including both of the positive parity
and the negative parity states.
We do not adjust the relative strength of
the spin-spin, the spin-orbit and the tensor term.
The result is reasonably consistent with the
experiments.
Thus, we can conclude that the flavor-singlet interaction
plays an essential role
in reducing the strong spin-orbit force in
the excited baryons to the observed strength.

In ref.\ \cite{IK78,Is92,KI80},
the tensor part of OGE was introduced
so as to give an appropriate decay rates.
One of their examples is the relative strength of the
$\pi$N decay from two $\Delta$($\FRAC{5}{2}^+$):
$\Delta$(1905) and $\Delta$(2000).
We estimate the ratio of the decay matrix elements for those states
using the transition operator defined in \cite{KI80},
where they assumed the pointlike pion is emitted from single quark.
The calculated matrix element for the
higher $\Delta$($\FRAC{5}{2}^+$)
is found by about 30\% smaller
than that of the lower state.
The experimental partial decay width
of the lower energy state to the
$\pi$N channel is 32 to 39 MeV;
two experiments are reported
for the decay width from the higher energy state:
 5 and 28 MeV \cite{Data}.
It seems that the higher state
decays more weakly to $\pi$N than the lower state.
This decay-rate ratio for these states is
consistent with the experiments,
though it should be considered as a very rough estimate.

A possible flaw of our model as well as that in
ref.\ \cite{IK78,Is92,KI80} is $\Delta$($\FRAC{3}{2}^+$).
Two $\Delta$($\FRAC{3}{2}^+$) are seen experimentally:
$\Delta$(1600) and $\Delta$(1920);
both of them decay to the $\pi$N
channel rather strongly.
The lowest energy level of the predicted states is 1734 MeV.
Inclusion of the spin-orbit term and \III\ has made the state
lower by about 150 MeV,
but the level is still higher than it should be by about 100 MeV.
The estimated decay matrix elements from
both of the two lower states
to the  $\pi$N channel are large, but that from the highest level
is small.
The couplings to the baryon-meson channel,
such as N$\pi$, may be important
 \cite{Fu93}.
The experiments have a large error also for these states.
Further investigation  both in the
experiments and in the theories is necessary to clarify this problem.

\subsection{Two Nucleons}

The main purpose of this paper is to show the inclusion of the
instanton induced interaction
gives the channel specific cancellation of
the spin-orbit force between quarks.
In the symmetry consideration we show that
the spin-orbit force in the two-nucleon scattering
does not reduce much by introducing \III.
In this section we show that a realistic
quark cluster model including
\III\
can actually reproduce the two-nucleon scattering phase shift
for the triplet $P$-wave states.

The wave function is the same as those in ref.\
\cite{MYO84,TSY89,TO91}:
\begin{equation}
	\Psi = \calA_{q}\{\phi^{2}_{N}\chi(R)\} ~~.
\end{equation}
The notation is the same as that in the appendix,
except for $\chi(R)$, the relative wave function,
which is now to be solved.

The hamiltonian for the valence quarks
is eq.(\ref{eq:quarkmodel}), except that
we omit the tensor terms of OGE and \III\
because the tensor force
between the quarks is not dominant
in the two-nucleon system \cite{TSY89}.
We use the linear confinement potential for $V_{\rm conf}$
in eq.(\ref{eq:quarkmodel}):
\begin{equation}
	V_{\rm conf} = \sum_{i<j} \: a_{\rm conf}\; r_{ij} ~~.
\end{equation}
After integrating out the internal coordinates of the nucleons,
 we have the resonating group method equation:
\begin{eqnarray}
	\{ H_{q}+N^{1/2}V_{\rm EMEP}N^{1/2}-EN\}\chi  & = & 0
	\label{eq:RGMeq}
\end{eqnarray}
with $H_{q}$ is the hamiltonian kernel for $H_{quark}$,
 N is the normalization kernel.
The effective meson exchange potential, $V_{\rm EMEP}$,
is multiplied by $N^{1/2}$
because the potential $V_{\rm EMEP}$ should be added
to the equation in the Schr\"odinger form.
Here we take $V_{\rm EMEP}$ to have the central
and the tensor parts of
the one-pion exchange with
the form factor corresponding to the size parameter $b$,
and the gaussian-type central attraction \cite{TSY89}:
\begin{eqnarray}
	V_{\rm EMEP}(R) & = &
	(\tau\cdot\tau)(\sig\cdot\sig) V_{\pi}^{C}(R)+
	(\tau\cdot\tau) V_{\pi}^{T}(R)\;S_{12}+V_{g}(R)
	\label{eq:EMEP}
	\\
	V_{\pi}^{C}(R) & = &
	 {g^{2}_{\pi}\over 4\pi}{1\over 3}
	 \left({m_{\pi}\over2m_{N}}\right)^{2}
{\exp[-m_{\pi}R]\over R}
{1\over 2}\{ {\rm Erfc}(\alpha_{-})
- \exp[2m_{\pi}R]{\rm Erfc}(\alpha_{+})  \}
\\
	V_{\pi}^{T}(R) & = & {g^{2}_{\pi}\over 4\pi}
	{1\over 3} \left({m_{\pi}\over2m_{N}}\right)^{2}
{\exp[-m_{\pi}R]\over R} {1\over 2}
	\left[\left\{1+{3\over m_{\pi}R}+
	{3\over (m_{\pi}R)^{2}}\right\}
	   {\rm Erfc}(\alpha_{-}) \right.
\nonumber\\
&-& \left\{1-{3\over m_{\pi}R}
+{3\over (m_{\pi}R)^{2}}\right\}\exp[2m_{\pi}R]{\rm Erfc}(\alpha_{+})
\nonumber\\
&-&\left.\left\{1+{6\beta\over(m_{\pi}R)^{2}}\right\}
{m_{\pi}R\over \sqrt{\pi\beta^{3}}} \exp[-\alpha_{-}^{2}] \right]
	\\
\alpha_{\pm} & = & {m_{\pi} b \over \sqrt{3}}
\pm {\sqrt{3} R \over 2b}
\\
\beta &=& (m_{\pi} b)^{3}/3
\\
	V_{g}(R)&=&
	V_{\sigma}
	\exp\left[-\left({R\over r_{\sigma}+r_{a}}\right)^{2}\right]
	-\{V_{\sigma}+V_{\pi}^{C}(0)\}
	\exp\left[-\left({R\over r_{\sigma}-r_{a}}\right)^{2}\right] ~~.
\end{eqnarray}
The parameters $V_{\sigma}$ and $r_{\sigma}$ in $V_{\rm EMEP}$
are determined
by fitting the experimental phase shifts of
the triplet partial-odd wave states
with  fixed $r_{a}$  (table 1.)
 The coupling constant
$g^{2}_{\pi}/(4\pi)=13.7$ is taken from ref.\ \cite{SAID}.
The other parameters
in $H_{quark}$ are the same as in the previous section,
which are also listed in table 1.

To see the contribution from the spin-orbit term more clearly,
we recompile the phase shift as \cite{MYO84,TSY89}
\begin{eqnarray}
	\delta(^{3}P_{C}) & = &
	{1\over9}\delta(^{3}P_{0})+{1\over3}\delta(^{3}P_{1})
	+{5\over9}\delta(^{3}P_{2})
\\
	\delta(^{3}P_{T}) & = &
-{5\over36}\delta(^{3}P_{0})+{5\over24}\delta(^{3}P_{1})
-{5\over72}\delta(^{3}P_{2})
\\
	\delta(^{3}P_{LS}) & = &
-{1\over6}\delta(^{3}P_{0})-{1\over4}\delta(^{3}P_{1})
+{5\over12}\delta(^{3}P_{2}) ~~.
\end{eqnarray}

The calculated phase shifts are shown in fig.\ 3
together with those of the
energy-dependent phase shift analysis for the low-energy region,
VZ40, taken from SAID database \cite{SAID}.
The central part seems to require more sophisticated
effective meson exchange than the two-ranged gaussian
potential.
The one-pion exchange can give enough strength to
the tensor part of the two-nucleon system.
The spin-orbit force is reproduced by
the quark model well, even when \III\ is included by $\piii = 0.4$.
As seen in table 2, the cancellation occurs only
for the color-singlet
quark pairs, which play minor role
in the $P$-wave two-nucleon systems.
The \III\ spin-orbit part for the color-singlet pairs
has the same sign as that of OGE.
Thus, the reduction of the spin-orbit effect
by including \III\ is less than
$(1-\piii)$.

\section{Discussion and Summary}

We investigate the effects of the noncentral part
of the instanton induced interaction
on the excited nonstrange baryons and two nucleon systems.
It is found that \III\ spin-orbit force cancels
the OGE spin-orbit force in the excited baryons
while the spin-orbit force in the two nucleon systems
remains strong after the
inclusion of \III.

The other possible source of the noncentral part
is the confinement force.
Such a channel specific cancellation, however,
occurs because that \III\ only affects
the flavor-singlet quark pairs.
The two-body confinement force, which has
the factor $(\lam\cdot\lam)$,
shows similar channel dependence to OGE and cannot produce
the cancellation required here.

The meson-exchange also produces the noncentral parts.
Usually, the $\sigma$,
$\rho$ and $\omega$ mesons are considered
as the main source of the spin-orbit force between
the nucleons.
This picture, however, cannot be applied
to the quark systems in a straightforward way.
These mesons are not pointlike; one cannot safely assume
that they
interact directly to quarks.
Moreover, their couplings to the nucleons are determined
mainly from the two-nucleon scattering data empirically.
In a model which includes only meson and baryon degrees of freedom,
the $\omega$N coupling is usually taken to be strong so
as to produce the short-range
repulsion between two nucleons.
In a quark model the repulsion is explained
 by the quark Pauli-blocking effect, by OGE and by \III.
The $\sigma$-meson exchange, which produces
the intermediate attraction,
also stands for complicated modes such as two-pion exchange,
coupling to the
$\pi\Delta$ channel, and even the attraction from \III.
It is hard to determine the genuine coupling to the mesons.
Here we take a quark-meson hybrid picture;
the short-ranged properties are explained
by the nature of a quark model
and the long-ranged properties are explained
by the meson-exchange model.
The modes which can be presented by the mesons
are taken into account as
the meson clouds.
The $\rho$-meson cloud has similar dependence
to OGE because
$(\tau\cdot\tau)$ shows similar channel dependence
to $(\lam\cdot\lam)$;
it will not produce the channel-specific cancellation to OGE.
The spin-orbit force of the $\omega$ or $\sigma$ cloud
 may produce such a cancellation.
Their strength, however, becomes smaller in such a picture.
 The contribution from meson clouds is unlikely
 to produce large effects required here.

On the contrary, there is a clear evidence for
the existence of the interaction
between the flavor-singlet quarks in the meson mass spectrum,
i.e.\ $\eta'$-$\eta$
mass difference;
this effect is considered to come not from the meson exchange
but from  the instanton-light quark coupling,
whose role we investigate here.
It is natural to think that there are large effects from \III\
also on the
 properties of the baryons or baryon systems, one of which,
 we argue,
 is the cancellation in the spin-orbit and the tensor force.

 The spin-orbit and the
 tensor part of \III\ survives only
 for the color-singlet quark pairs
 in a relative-odd partial wave state or
 for the color-octet quark pairs in a relative-even
 partial wave state.
 There, the cancellation of the spin-orbit and tensor parts
 occurs between OGE and \III;
 which leads the particular cancellation required.
The relative strength of each term may change if we take
into the higher order relativistic effects account.
 The channel-specific cancellation is, however,
 explained based on the symmetry;
the overall nature will still be valid
with the change of the model parameters.

 This discussion based on the symmetry have to
 be reexamined when one considers
 the relativistic systems or the systems including strangeness.
 The estimate by  the MIT bag model indicates that
the major effect of the cancellation still exists
in the negative-parity baryons
\cite{Ta94}.
It is interesting to investigate the role of
the instanton induced interaction
in the excited baryons with the strangeness,
especially in the flavor-singlet states,
which will be presented in elsewhere.
\bigskip

This work was supported in part by Grant-in-Aid
for Encouragement of Young Scientists (A) (No.\ 07740206),
for General Scientific Research (No.\ 04804012)
and for Scientific Research on Priority Areas (No.\ 05243102)
from the Ministry of Education, Science, Sports and Culture.

\newpage
\section*{
Appendix A: Matrix Elements by the harmonic oscillator wave
functions}
Here we evaluate each operator defined by eq.\ (\ref{eq:projop}).

\subsection{Single Baryons}
\subsubsection{Wave functions}

\noindent{\it a) $S$-wave baryons}

There are flavor-decuplet states with $J^P=\FRAC{3}{2}^+$
and flavor-octet
states with $\FRAC{1}{2}^+$.
 By writing the orbital angular momentum, $L$, and
intrinsic spin, S, explicitly with the dimension
in the flavor space, {\bf D}$^{F}$,
and the dimension in the spin-flavor space,  D$^{SF}$,
as $|{\bf D}^{F};D^{SF}(LS)J^{P}\ket$, they are represented as:
\begin{eqnarray}
|1\ket \equiv \left|{\bf 10};56\left(0\half(3)\right)\half(3)^{+}
\right\rangle
& = & |[1^3]^C\ket  |[3]^O \ket  |[3]^F \ket |[3]^S \ket
\label{eq:31}
\\
|2\ket \equiv \left|{\bf 8};56\left(0\half(1)\right)\half(1)^{+}
\right\rangle
& = & |[1^3]^C\ket  |[3]^O \ket
 \{|[21]^F \ket  |[21]^S \ket\}_{[3]} ~~.
\end{eqnarray}

\noindent{\it b) $P$-wave baryons}

There are flavor-decuplet states with $J^P=\FRAC{1}{2}^-$,
$\FRAC{3}{2}^-$ and flavor-octet
states with $J^P=(\FRAC{1}{2}^-)^{2}$, $(\FRAC{3}{2}^-)^{2}$,
$\FRAC{5}{2}^-$.  For a
future use,  we listed flavor-singlet states with
$\FRAC{1}{2}^-$, $\FRAC{3}{2}^-$
for the strangeness $-1$ systems.  In the same representation above,
they are:
\begin{eqnarray}
|3\ket \equiv \left|{\bf 10};70\left(1\half(1)\right)J^{-}
\right\rangle
& = &
|[1^3]^C\ket \{ |[21]^O \ket |[3]^F \ket |[21]^S \ket \}_{[3]}
\\
|4\ket \equiv \left|{\bf 8};70\left(1\half(3)\right)J^{-}
\right\rangle
& = &
|[1^3]^C\ket \{ |[21]^O \ket |[21]^F \ket  \}_{[3]} |[3]^S \ket
\\
|5\ket \equiv \left|{\bf 8};70\left(1\half(1)\right)J^{-}
\right\rangle
& = &
|[1^3]^C\ket \{ |[21]^O \ket |[21]^F \ket |[21]^S \ket \}_{[3]}
\\
|6\ket \equiv \left|{\bf 1};70\left(1\half(1)\right)J^{-}
 \right\rangle
& = &
|[1^3]^C\ket \{ |[21]^O \ket |[1^3]^F \ket |[21]^S \ket \}_{[1^3]}~~.
\end{eqnarray}

\noindent{\it c) $N=2$ positive parity baryons}

There are flavor decuplet states with
$J^P=(\FRAC{1}{2}^+)^{2}$, $(\FRAC{3}{2}^+)^{3}$,
$(\FRAC{5}{2}^+)^{2}$, and $\FRAC{7}{2}^+$,
flavor octet states with
 $(\FRAC{1}{2}^+)^{4}$, $(\FRAC{3}{2}^+)^{5}$,
$(\FRAC{5}{2}^+)^{3}$, and $\FRAC{7}{2}^+$,
and flavor singlet states with
 $(\FRAC{1}{2}^+)^{2}$, $(\FRAC{3}{2}^+)^{2}$, and
$(\FRAC{5}{2}^+)^{2}$:
\begin{eqnarray}
|7_L\ket \equiv \left|{\bf 10};56\left(L\half(3)\right)J^{+}
\right\rangle
& = & |[1^3]^C\ket  |[3]^O \ket |[3]^F \ket |[3]^S \ket
\\
|8_L\ket \equiv \left|{\bf 10};70\left(L\half(1)\right)J^{+}
\right\rangle
& = &
|[1^3]^C\ket  \{ |[21]^O \ket |[3]^F \ket |[21]^S \ket \}_{[3]}
\\
|9_L\ket \equiv \left|{\bf 8};70\left(L\half(3)\right)J^{+}
\right\rangle
& = &
|[1^3]^C\ket \{|[21]^O \ket |[21]^F \ket \}_{[3]} |[3]^S \ket
\\
|10_L\ket \equiv \left|{\bf 8};56\left(L\half(1)\right)J^{+}
\right\rangle
& = &
|[1^3]^C\ket  |3]^O \ket \{|[21]^F \ket |[21]^S \ket  \}_{[3]}
\\
|11_L\ket \equiv \left|{\bf 8};70\left(L\half(1)\right)J^{+}
\right\rangle
& = &
|[1^3]^C\ket \{ |[21]^O \ket |[21]^F \ket |[21]^S \ket \}_{[3]}
\\
|13_L\ket \equiv \left|{\bf 1};70\left(L\half(1)\right)J^{+}
\right\rangle
& = &
|[1^3]^C\ket  \{ |[21]^O \ket |[1^3]^F \ket |[21]^S \ket \}_{[1^3]}
\end{eqnarray}
with $L$ = 0 and 2, and
\begin{eqnarray}
|12\ket \equiv \left|{\bf 8};20\left(1\half(1)\right)J^{+}
\right\rangle
& = &
|[1^3]^C\ket |1^3]^O \ket \{|[21]^F \ket |[21]^S \ket  \}_{[1^3]}
\\
|14\ket \equiv \left|{\bf 1};20\left(1\half(3)\right)J^{+}
\right\rangle
& = &
|[1^3]^C\ket  |[1^3]^O  \ket  |[1^3]^F \ket |[3]^S \ket
\label{eq:44}
\end{eqnarray}
for $L$ = 1.

Here
\begin{eqnarray}
	\{|[21]^\alpha \ket  |[21]^\beta \ket\}_{[3]}
	&=& {1\over \rttwo } \left\{
 |[21]^\alpha \MS \ket |[21]^\beta \MS \ket
 +|[21]^\alpha \MA \ket |[21]^\beta \MA \ket
\right\}
\\
	\{|[21]^\alpha \ket  |[21]^\beta \ket\}_{[1^3]}
	&=& {1\over \rttwo} \left\{
|[21]^\alpha \MS \ket |[21]^\beta \MA \ket
- |[21]^\alpha \MA \ket |[21]^\beta \MS \ket
\right\}
\\
\{|[21]^\alpha \ket |[21]^\beta \ket |[21]^\gamma \ket\}_{[3]}
		&=& {1\over 2} \left\{
-  |[21]^\alpha \MS \ket |[21]^\beta \MS \ket |[21]^\gamma \MS \ket
+  |[21]^\alpha \MS \ket |[21]^\beta \MA \ket |[21]^\gamma \MA \ket
\right.
\nonumber\\
&+& \left. \!|[21]^\alpha \MA \ket |[21]^\beta \MS \ket |[21]^\gamma
\MA \ket
\!+\!
|[21]^\alpha \MA \ket |[21]^\beta \MA \ket |[21]^\gamma \MS \ket
\right\} ~.
\end{eqnarray}
The orbital wave function in the
coordinate space can be written by Jacobi's coordinates,
$\mbf(\xi) =
(\mbf(r)_1-\mbf(r)_2)/\sqrt{2}$,
$\mbf(\eta) = (\mbf(r)_1+\mbf(r)_2-2\mbf(r)_3)/\sqrt{6}$ and
$\mbf(R)_G=(\mbf(r)_1+\mbf(r)_2+\mbf(r)_3)/\sqrt{3}$.
When we write $ |NL[f]\ket $, they are
\begin{eqnarray}
|00[3]\ket & = & \psi_{0s}(\mbf(\xi))\psi_{0s}(\mbf(\eta))
\\
|01[21]\MS\ket & = & \psi_{0s}(\mbf(\xi))\psi_{0p}(\mbf(\eta))
\\
|01[21]\MA\ket & = & \psi_{0p}(\mbf(\xi))\psi_{0s}(\mbf(\eta))
\\
|01[1^{3}]\ket & = & [\psi_{0p}(\mbf(\xi))
\times \psi_{0p}(\mbf(\eta))]^{1}
\\
|02[3]\ket & = & {1\over \sqrt{2}}
\{ \psi_{0d}(\mbf(\xi))\psi_{0s}(\mbf(\eta))
+\psi_{0s}(\mbf(\xi))\psi_{0d}(\mbf(\eta)) \}
\\
|02[21]\MS\ket & = & {1\over \sqrt{2}}
\{ \psi_{0d}(\mbf(\xi))\psi_{0s}(\mbf(\eta))
-\psi_{0s}(\mbf(\xi))\psi_{0d}(\mbf(\eta)) \}
\\
|02[21]\MA\ket & = & [ \psi_{0p}(\mbf(\xi))
\times \psi_{0p}(\mbf(\eta))]^{2}
\\
|10[3]\ket & = & {1\over \sqrt{2}}
\{ \psi_{1s}(\mbf(\xi))\psi_{0s}(\mbf(\eta))
+\psi_{0s}(\mbf(\xi))\psi_{1s}(\mbf(\eta)) \}
\\
|10[21]\MS\ket & = & {1\over \sqrt{2}}
\{ \psi_{1s}(\mbf(\xi))\psi_{0s}(\mbf(\eta))
-\psi_{0s}(\mbf(\xi))\psi_{1s}(\mbf(\eta)) \}
\\
|10[21]\MA\ket & = & [ \psi_{0p}(\mbf(\xi))
\times \psi_{0p}(\mbf(\eta))]^{0} ~~,
\end{eqnarray}
where $[a_{l}\times b_{l'}]^{L}_{M}
= \sum (lml'm'|LM) a_{lm}b_{l'm'}$.

The flavor part is the same as in ref.\ \cite{Is92}.
For the proton, it is
\begin{eqnarray}
    |[3]\ket	 & = & (duu+udu+uud)
  \\
	|[21]\MS\ket & = & {1\over\sqrt{6}}(-udu-duu+2uud)
  \\
	|[21]\MA\ket & = & {1\over\sqrt{2}}(udu-duu) ~~.
\end{eqnarray}

\subsubsection{Matrix Elements}
The operators we consider can be written as
\begin{eqnarray}
	\calO & = & \sum_{(i<j)} \left\{
	  \calO_c ^{C}  (\calS^F\calO_c^O+\calA^F{\barcalO}_c^O)~1
+ \calO_{ss}^{C}(\calS^F\calO_{ss}^O
+\calA^F{\barcalO}_{ss}^O)~\sigma\cdot\sigma
\right.\nonumber\\
&+& \left.
 \calO_{ls}^{C}(\calS^F\calO_{ls}^O
 +\calA^F{\barcalO}_{ls}^O)~L\cdot S
+ \calO_{t}^{C}(\calS^F\calO_{t}^O
+\calA^F{\barcalO}_{t}^O)~S_{12}\right\}.
\label{eq:symO}
\end{eqnarray}
Then, the matrix element
\begin{eqnarray}
	\calO(n'_{L'},n_{L}) & \equiv &
	 {1\over \sqrt{2J+1}}\bra n'(LS')J || \calO ||n(LS)J \ket
\end{eqnarray}
 is reduced to sum of the two-body
matrix elements.  Here $n$ is the number expressing the states
defined  by the eqs. (\ref{eq:31}) -- (\ref{eq:44}).

\noindent{\it a) Flavor-decuplet positive parity baryons}

\begin{eqnarray}
\calO(1,1)
& = & A_{0s}+ \Del_{0s}
\label{eq:mateleS10}\\
\calO(7_0,7_0)
& = &
{1\over 2} \leftw A_{0s}+A_{1s}+\Del_{0s}+\Del_{1s}\rightw
\label{eq:matele1S}
\\
\calO(7_{0},7_{2})
& = & {1\over 2}C_{t}(0\half(3)2\half(3)J)\sqrt{5}\;
\theta_{1s-0d}
\\
\calO(7_{0},8_{2})
& = & {1\over2\sqrt{2}} C_{t}(0\half(3)2\half(1)J)\sqrt{10}\;
\theta_{1s-0d}
\\
\calO(8_{0},8_{0})
& = &
{1\over 4} \leftw A_{0s}+A_{1s}+2A_{0p}
+\Del_{0s}+\Del_{1s}-6\Del_{0p}\rightw
\\
\calO(8_{0},7_{2})
& = &
{1\over 2\sqrt{2}} C_{t}(0\half(1)2\half(3)J)(-\sqrt{10})\;
 \theta_{1s-0d}
\\
\calO(7_{2},7_{2})
& = &
{1\over 2} \leftw A_{0s}+A_{0d}+\Del_{0s}+\Del_{0d}
\nonumber\\
&+& C_{ls}(2\half(3)2\half(3)J)(-\sqrt{3\over 2})\;
\chi_{0d}
+ C_{t}(2\half(3)2\half(3)J)(-\sqrt{10\over 7})\;
\theta_{0d}\rightw
\\
\calO(7_{2},8_{2}) &=&
{1\over 2\sqrt{2}}\leftw
 C_{ls}(2\half(3)2\half(1)J)(-\sqrt{3\over 5})\;
 \chi_{0d}
+C_{t}(2\half(3)2\half(1)J)(-\sqrt{20\over 7})\;
\theta_{0d}\rightw
\\
\calO(8_{2},8_{2})
& = &
{1\over 4} \leftw A_{0s}+A_{0d}+2A_{0p}
+\Del_{0s}+\Del_{0d}-6\Del_{0p}
\nonumber\\
&+& C_{ls}(2\half(1)2\half(1)J)(-\sqrt{12\over 5})\;
 \chi_{0d}\rightw
\\
\end{eqnarray}

\noindent{\it b) Flavor-octet positive parity baryons}

\begin{eqnarray}
\calO(2,2)
& = & {1\over2} \leftw A_{0s}+ \Del_{0s}+\barA_{0s}
-3\barDel_{0s}\rightw
\\
\calO(9_{0},9_{0})
& = &
{1\over 4} \leftw A_{0s}+A_{1s}+2 \barA_{0p}
+\Del_{0s}+\Del_{1s}+2\barDel_{0p}\rightw
\\
\calO(9_{0},12)
& = & -{1\over 2} C_{ls}(0\half(3)1\half(1)J)
\sqrt{1\over 3}\barchi_{0p}
\\
\calO(9_{0},9_{2})
& = &
{1\over 4}
 C_{t}(0\half(3)2\half(3)J)\leftw (-\sqrt{2})2\bartheta_{0p}
+\sqrt{5}\; \theta_{1s-0d}\rightw
\\
\calO(9_{0},10_{2})
& = &
{1\over 4}
 C_{t}(0\half(3)2\half(1)J)\sqrt{10}\; \theta_{1s-0d}
\\
\calO(9_{0},11_{2})
& = &
{1\over 4\sqrt{2}}
 C_{t}(0\half(3)2\half(1)J)\leftw (-\sqrt{4})2 \bartheta_{0p}
-\sqrt{10} \; \theta_{1s-0d}\rightw
\\
\calO(10_{0},10_{0})
& = &
{1\over 4}
\leftw A_{0s}+A_{1s}+\barA_{0s}+\barA_{1s}
+\Del_{0s}+\Del_{1s}-3\barDel_{0s}-3\barDel_{1s}\rightw
\\
\calO(10_{0},11_{0})
& = &
{1\over 4\sqrt{2}}
\leftw A_{0s}-A_{1s}-\barA_{0s}+\barA_{1s}+
\Del_{0s}-\Del_{1s}+3\barDel_{0s}-3\barDel_{1s}\rightw
\\
\calO(10_{0},9_{2})
& = &
{1\over 4}
 C_{t}(0\half(1)2\half(3)J)(-\sqrt{10})\; \theta_{1s-0d}
\\
\calO(11_{0},11_{0})
& = &
{1\over 8}
\leftw A_{0s}+A_{1s}+2A_{0p}+\barA_{0s}+\barA_{1s}+2\barA_{0p}
\nonumber\\
&&+\Del_{0s}+\Del_{1s}-6\Del_{0p}-3\barDel_{0s}-3\barDel_{1s}
+2\barDel_{0p}\rightw
\\
\calO(11_{0},12)
& = &
-{1\over 2\sqrt{2}}
 C_{ls}(0\half(1)1\half(1)J)\sqrt{4\over 3}\barchi_{0p}
\\
\calO(11_{0},9_{2})
& = &
{1\over 4\sqrt{2}}
C_{t}(0\half(1)2\half(3)J)\leftw \sqrt{4}\: 2 \bartheta_{0p}
 -(-\sqrt{10})\bartheta_{1s-0d}\rightw
\\
\calO(9_{2},9_{2})
& = &
{1\over 4} \leftw A_{0s}+A_{0d}+2 \barA_{0p}+\Del_{0s}+\Del_{0d}
+2\barDel_{0p}
\nonumber\\
&+& C_{ls}(2\half(3)2\half(3)J)((-\sqrt{3\over8})2\barchi_{0p}
+(-\sqrt{3\over2})\; \chi_{0d})
\nonumber\\
&+&C_{t}(2\half(3)2\half(3)J)((-\sqrt{7\over 10})2\bartheta_{0p}
+(-\sqrt{10\over7})\; \theta_{0d})
\rightw
\\
\calO(9_2,10_2) &=&
{1\over 4}  \leftw C_{ls}(2\half(3)2\half(1)J)(-\sqrt{3\over 5})\;
 \chi_{0d}
+C_{t}(2\half(3)2\half(1)J)(-\sqrt{20\over 7})\; \theta_{0d}
\rightw
\\
\calO(9_2,11_2) &=&
{1\over 4\sqrt{2}} \leftw C_{ls}(2\half(3)2\half(1)J)
((-\sqrt{3\over 20})2\barchi_{0p}-(-\sqrt{3\over 5})\;
 \chi_{0d})
\nonumber\\
&+&C_{t}(2\half(3)2\half(1)J)((-\sqrt{7\over 5})2
\bartheta_{0p}-(-\sqrt{20\over 7})\; \theta_{0d})\rightw
\\
\calO(9_2,12) &=&
-{1\over 2}  \leftw C_{ls}(2\half(3)1\half(1)J)
(-\sqrt{1\over 12})\barchi_{0p}
+C_{t}(2\half(3)1\half(1)J) (-\sqrt{3})\bartheta_{0p}\rightw
\\
\calO(10_2,10_2)
& = &
{1\over 4}
\leftw A_{0s}+A_{0d}+\barA_{0s}+\barA_{0d}+\Del_{0s}
+\Del_{0d}-3\barDel_{0s}-3\barDel_{0d}
\nonumber\\
&+& C_{ls}(2\half(1)2\half(1)J)(-\sqrt{12\over 5})\;
 \chi_{0d}\rightw
\\
\calO(10_2,11_2)
& = &
{1\over 4\sqrt{2}}
\leftw A_{0s}-A_{0d}-\barA_{0s}+\barA_{0d}+
\Del_{0s}-\Del_{0d}+3\barDel_{0s}-3\barDel_{0d}
\nonumber\\
&-&  C_{ls}(2\half(1)2\half(1)J)(-\sqrt{12\over 5})\;
\chi_{0d}\rightw
\\
\calO(11_2,11_2)
& = &
{1\over 8}
\leftw A_{0s}+A_{0d}+2A_{0p}+\barA_{0s}+\barA_{0d}+2\barA_{0p}
\nonumber\\
&&~~ + \Del_{0s}+\Del_{0d}-6\Del_{0p}-3\barDel_{0s}-3\barDel_{0d}
+2\barDel_{0p}
\nonumber\\
&+& C_{ls}(2\half(1)2\half(1)J) ((-\sqrt{3\over 5})2\barchi_{0p})
+(-\sqrt{12\over 5})\; \chi_{0d}\rightw
\\
\calO(11_2,12) &=& -{1\over 2\sqrt{2}}
C_{ls}(2\half(1)1\half(1)J) (-\sqrt{1\over 3})\barchi_{0p}
\\
\calO(12,12)
& = &
{1\over 2} \leftw A_{0p}+\barA_{0p}-3\Del_{0p}+\barDel_{0p}
+ C_{ls}(1\half(1)1\half(1)J)(-\sqrt{1\over 3})\barchi_{0p}\rightw
\\
\end{eqnarray}

\noindent{\it c) Flavor-singlet positive parity baryons}

\begin{eqnarray}
\calO(13_{0},13_{0}) & = &{1\over 4} \leftw 2\barA_{0p}+\barA_{0s}
+\barA_{1s}+2\barDel_{0p}
-3\barDel_{0s}-3\barDel_{1s}\rightw
	\\
\calO(13_{0},14) &=& -{1\over\sqrt{2}}
 C_{ls}(0\half(1)1\half(3)J)(-\sqrt{1\over 3})\barchi_{0p}
\\
\calO(13_{2},13_{2}) & = &{1\over 4} \leftw 2\barA_{0p}+\barA_{0s}
+\barA_{0d}+2\barDel_{0p}
-3\barDel_{0s}-3\barDel_{0d}
\nonumber\\
&+& C_{ls}(2\half(1)2\half(1)J)(-\sqrt{3\over 5})2\barchi_{0p}
\rightw
	\\
\calO(13_{2},14) &=& -{1\over\sqrt{2}}\leftw
 C_{ls}(2\half(1)1\half(3)J)\sqrt{1\over 12}\barchi_{0p}
+C_{t}(2\half(1)1\half(3)J)\sqrt{3}\bartheta_{0p}
 \rightw
\\
\calO(14,14) & = & \barA_{0p}+\barDel_{0p}
\nonumber\\
&+& C_{ls}(1\half(3)1\half(3)J)(-\sqrt{5\over 24})\barchi_{0p}
+ C_{t}(1\half(3)1\half(3)J)\sqrt{5\over 6}\bartheta_{0p}
\end{eqnarray}

\noindent{\it d) Negative parity baryons}

\begin{eqnarray}
\calO(3,3)
& = &{1\over2} \leftw A_{0s}+A_{0p}+ \Del_{0s}-3 \Del_{0p}\rightw
\label{eq:mateleP10}\\
\calO(4,4)
& = &{1\over2} \leftw A_{0s}+\Del_{0s}+\barA_{0p}+\barDel_{0p}
\nonumber\\
&+& C_{ls}(1\half(3)1\half(3)J)(-\sqrt{5\over 6})\barchi_{0p}
+ C_{t}(1\half(3)1\half(3)J)(-\sqrt{10\over 3}){\bartheta}_{0p}
\rightw
	\\
\calO(4,5)
& = & {1\over 2\sqrt{2}}\leftw C_{ls}(1\half(3)1\half(1)J)
 (-\sqrt{1\over 3}) \barchi_{0p}
+ C_{t}(1\half(3)1\half(1)J) (-\sqrt{20\over 3}) {\bartheta}_{0p}
\rightw
\\
\calO(5,5)
& = &{1\over 4} \leftw
A_{0s}+A_{0p} +\Del_{0s}-3\Del_{0p}+\barA_{0s}+\barA_{0p}
-3\barDel_{0s}+\barDel_{0p}
\nonumber\\
&+& C_{ls}(1\half(1)1\half(1)J) (-\sqrt{4\over 3}) \barchi_{0p}
\rightw
\\
\calO(6,6)
& = &{1\over2} \leftw \barA_{0s}+\barA_{0p}-3\barDel_{0s}
+\barDel_{0p}
+ C_{ls}(1\half(1)1\half(1)J) (-\sqrt{4\over 3}) \barchi_{0p}
\rightw ~~.
\label{eq:mateleP1}
\end{eqnarray}
$C_{ls}(L'S'LSJ)$ and $C_t(L'S'LSJ)$ are defined as
\begin{eqnarray}
	C_{\alpha}(L'S'LSJ) & = &
	 \sqrt{(2J+1)(2L'+1)(2L+1)(2S'+1)(2S+1)}
	\ninej(L',S',J,\lambda,\lambda,0,L,S,J)
\end{eqnarray}
with $\lambda$ = 1 and 2 for $\alpha$ = {\it ls} and {\it t},
 respectively.
The radial integrations are:
\begin{eqnarray}
	A_{nl} & = & 3\bra nl|{\calO_c^O}_{(ij)=(12)}
	|nl\ket \bra {\calO_{c}^{C}}_{(ij)=(12)} \ket
{} \\
	\Del_{nl} & = & 3\bra nl|{\calO_{ss}^O}_{12}
	|nl\ket \bra \calO_{ss}^{C} \ket
{} \\
	\chi_{nl} & = &3 \bra nl|{\calO_{ls}^O}_{12}
	|nl\ket \bra \calO_{ls}^{C} \ket
{} \\
	\theta_{nl} & = & 3\bra nl|{\calO_{t}^O}_{12}
	|nl\ket \bra \calO_{t}^{C} \ket
{} \\
	\theta_{n'l'-nl} & = & 3\bra n'l'|{\calO_{t}^O}_{12}
	|nl\ket \bra \calO_{t}^{C} \ket ~~.
{}
\end{eqnarray}
Those with bars are similarly defined.
{}From eqs.(\ref{eq:mateleS10})--(\ref{eq:mateleP1}),
one can actually see
only the flavor-antisymmetric $0p$ pairs
and the flavor-symmetric $0d$ pairs are relevant
for the noncentral
part in the single baryons.

 For OGE, the terms in  eq.\ (\ref{eq:symO}) are
	\begin{eqnarray}
	\calO^O_{ss} = \barcalO^O_{ss} & = &
	 - 4\pi\alphas {1\over 6m^2} \delta(\mbfr)
		  \\
	\calO^O_{ls} = \barcalO^O_{ls} & = &
	 - 4\pi\alphas {3\over 2m^2}{1\over 4\pi r^3}
		  \\
	\calO^O_{t} = \barcalO^O_{t} & = &
	- 4\pi\alphas {1\over 4m^2}{1\over 4\pi r^3}
	\end{eqnarray}
with
$$
\calO^C_\alpha = {\lambda\cdot\lambda \over 4}.
$$
Therefore, the two-body matrix elements become:
\begin{eqnarray}
	\Del_{0s} = \barDel_{0s} & = & {1\over 2}\mu_{\rm OGE}
\\
	\Del_{1s} = \barDel_{1s} & = & {3\over 4}\mu_{\rm OGE}
\\
	\Del_{nl} = \barDel_{nl} & = & 0 ~~(l>0)
\\
	\barchi_{0p} & = & {3\over 2}\mu_{\rm OGE}
\label{eq:barchi0pOGE}
\\
	\chi_{0d} & = & {3\over 5}\mu_{\rm OGE}
\\
	\bartheta_{0p} & = &  {1\over 4}\mu_{\rm OGE}
\label{eq:bartheta0pOGE}
\\
	\theta_{0d} & = &  {1\over 10}\mu_{\rm OGE}
\\
	\theta_{0d-1s} & = &  \sqrt{1\over 160}\mu_{\rm OGE}
\end{eqnarray}
with
$$
\mu_{\rm OGE} =
\half(1) (A_{0s}+\Del_{0s}-\barA_{0s}+3\barDel_{0s})=
 \alphas {4\over 3} {1\over\sqrt{2\pi}} {1\over m^2b^3},
$$
which corresponds to the $S$-wave N-$\Delta$ mass difference
(= 293 MeV).

As for \III, all operators without a bar vanish.  The
flavor-singlet operators are
	\begin{eqnarray}
	\barcalO^O_{ss} & = &
	 - \Vzerotwo {3\over 16} \delta(\mbfr)
		  \\
	\barcalO^O_{ls} & = &
	 \Vzerotwo  {9\over 4m^2} {\delta(r) \over 4\pi r^{4}}
		  \\
	\barcalO^O_{t} & = &
	  \Vzerotwo  {5\over 4m^2} {\delta(r) \over 4\pi r^{4}}
	\end{eqnarray}
with
\begin{eqnarray}
\calO^C_{ss} =
\calO^C_{ls} &=& 1 \\
\calO^C_t &=& 1 - {3\over4}{\lambda\cdot\lambda \over 4} ~~.
\end{eqnarray}
Thus we obtain the two-body matrix elements for \III\ as
\begin{eqnarray}
	\barDel_{0s} & = & {1\over 4}\mu_{\rm III}
\\
	\barDel_{1s} & = & {3\over 8}\mu_{\rm III}
\\
	 \barDel_{nl} & = & 0 ~~(l>0)
\\
	\barchi_{0p} & = & -{1\over m^{2}b^{2}} \mu_{\rm III}
\label{eq:barchi0pIII}
\\
	\barchi_{nl} & = & 0 ~~(l>1)
\\
	\bartheta_{0p} & = &  -{5\over 6}{1\over m^{2}b^{2}} \mu_{\rm III}
\label{eq:bartheta0pIII}
\\
\bartheta_{nl} & = &  0 ~~(l>2)
\end{eqnarray}
with
$$
\mu_{\rm III} =
- {9\over 4}  \Vzerotwo {1\over\sqrt{2\pi b^{2}}^{3}} ~~.
$$

The mass of baryons are obtained from
eq.(\ref{eq:mateleS10})-(\ref{eq:mateleP1}),
 by diagonarizing them if
necessary, and adding the central part.
\subsection{Two Nucleons}

Here we consider a six-quark system $(0s)^{5}(0p)$,
with quantum number of
two nucleons with relative partial $P$-wave.
 The wave function is
\begin{equation}
	\Psi = \calA_{q}\Phi \equiv
	\calA_{q}\{\phi^{2}_{N}\psi_{0p}(R)\}	 ~~,
\end{equation}
where $\calA_{q}$ is an antisymmetrizer with respect
to all six quarks,
$\phi_{N}$ is the nucleon wave function defined
by the previous section,
and the $\psi_{0p}(R)$ is a $P$-wave harmonic oscillator
with size parameter
$b$ and $R=(r_{1}+r_{2}+r_{3}-r_{4}-r_{5}-r_{6})/\sqrt{6}$.

Noncentral operators relevant to this state
are orbitally-antisymmetric two terms in eq.(\ref{eq:projop}).
The expectation value by the above
state can written as

\begin{eqnarray}
	\bra \Psi |\calO | \Psi\ket & = &
\bra \Phi |\sum_{i<j}(\barcalO^A_{ij} + \calO^S_{ij}
)\calA_{q}|\Phi\ket
/
\bra \Phi |\calA_{q}|\Phi\ket
 \\
	 & = &
\bra \Phi |\sum_{i<j}(\barcalO^A_{ij} +
\calO^S_{ij})(1-9P_{36})|\Phi\ket
/
\bra \Phi |(1-9P_{36})|\Phi\ket  ~~,
\label{eq:143}
\end{eqnarray}
where $P_{36}\equiv P_{36}^{c}P_{36}^{f}P_{36}^{s}P_{36}^{o}$
is the exchange operator for quark $i$, and $j$ in the color,
flavor,
spin and orbital space.
The numerator is
\begin{eqnarray}
&   & 	9\bra \Phi| (\barcalO^A_{36} + \calO^S_{36})
 |\Phi\ket
-	9\leftw 4 \bra \Phi| (\barcalO^A_{14} + \calO^S_{14})P_{36}
|\Phi\ket
+ \bra \Phi | (\barcalO^A_{36} + \calO^S_{36})P_{36}
 |\Phi\ket \rightw
\nonumber\\
& = & 	9\leftw 2\bra\Phi| (\barcalO^A_{36} + \calO^S_{36})
|\Phi\ket
- 	4 \bra \Phi | (\barcalO^A_{14} + \calO^S_{14})P_{36}
|\Phi\ket\rightw
\nonumber\\
& = & 	18\leftw \bra \Phi|\barcalO^A_{36}|\Phi\ket
+\bra \Phi |\calO^S_{36}|\Phi\ket - 2\bra \Phi|\calO^S_{14}P_{36
}|\Phi\ket\rightw
\label{eq:2Nmatele}
\end{eqnarray}
Here we use
\begin{eqnarray}
	\calO_{16}P_{36} & = & P_{36}\calO_{13}
\\
	\bra\Phi|\calO_{13}|\Phi\ket & = & \bra\Phi|\calO_{13} P_{36}
	|\Phi\ket = 0
\\
	\bra\Phi|\calO_{36} P_{36}|\Phi\ket & = & -\bra\Phi|\calO_{36}
	|\Phi\ket ~~.
\end{eqnarray}
For the spin-orbit part,
 $\calO_{ij} = f(r_{ij})L_{ij}\cdot S_{ij}$,
 each term in eq.(\ref{eq:2Nmatele}) can be evaluated as
\begin{eqnarray}
	\bra \Phi | \calO_{36} | \Phi\ket & = & {1\over16}\bra
 f(r_{36})L_{36}\cdot S_{36}(1\mp P_{36}^{c})
 (1\mp P_{36}^{f})(1+P_{36}^{s})(1-P_{36}^{o}) \ket
	  \\
	 & = & {1\over 2}(1\mp \bra P_{36}^{c}\ket)
	  {1\over 2}(\bra f(r_{36})L_{36} (1\!-\!P_{36}^{o}) \ket)
	 {1\over 4}\bra S_{36}(1+\!P_{36}^{s}\mp \! P_{36}^{f}
	 \mp \! P_{36}^{sf}) \ket
	  \\
	\bra \Phi | \calO_{14}P_{36} | \Phi\ket & = &
	{1\over16}\bra
 f(r_{14})L_{14}\cdot S_{14}(1\mp P_{14}^{c})(1\mp P_{14}^{f})
 (1+P_{14}^{s})(1-P_{14}^{o})P_{36} \ket
	  \\
	 & = & {1\over 2}(\bra P_{36}^{c}\ket\mp\bra
	  P_{14}^{c}P_{36}^{c}\ket)
	 {1\over 2}(\bra f(r_{14})L_{14}(1-P_{14}^{o})P_{36}^{o} \ket)
	  \nonumber\\&&
\times {1\over 4}\bra S_{14}(1+P_{14}^{s}\mp P_{14}^{f}
\mp P_{14}^{sf})P_{36}^{sf}\ket ~~.
\end{eqnarray}
The $\mp$ reads $-$ for $\barcalO^{A}$ and $+$ for $\calO^{S}$.
The color part is  $\bra P_{36}^{c}\ket = \bra P_{14}^{c}P_{36}^{c}
\ket = {1\over3}$.
The spin-flavor part can be calculated directly:
\begin{eqnarray}
	\bra p^\uparrow p^\uparrow | S_{36}(1+P_{36}^{s}\mp P_{36}^{f}
	\mp P_{36}^{sf}) |p^\uparrow p^\uparrow\ket & = &
	{1\over4}\leftw {1\over3}+{1\over3}\mp{7\over27}\mp{7\over27}
	\rightw = \left\{ \matrix{ {1\over27} \cr {8\over27}}\right.
	  \\
	\bra p^\uparrow p^\uparrow | S_{14}(1+P_{14}^{s}\mp P_{14}^{f}
	\mp P_{14}^{sf})  P_{36}^{sf} |p^\uparrow p^\uparrow\ket & = &
	{1\over4}\leftw {5\over81}+{5\over81}\mp{5\over81}\mp{5\over81}
	\rightw = \left\{\matrix{ 0 \cr {5\over81}}\right. .
\end{eqnarray}
$\bra f(r_{ij})L_{ij}(1-P_{36}^{o}) \ket = {2 \over 9}
\barchi^{A}_{0p}$ (or ${2 \over 9} \chi^{S}_{0p}$)
for $\barcalO^{A}$ (or $\calO^{S}$) and
$\bra f(r_{ij})L_{ij}(1-P_{36}^{o}) \ket = {2 \over 9}
\chi^{S}_{0p}$.

The denominator of eq.(\ref{eq:143}) is ${50\over 81}$
\cite{MYO84,TSY89}.  Thus
the matrix element eq.(\ref{eq:2Nmatele}) is
\begin{eqnarray}
		\bra \Psi |\calO | \Psi\ket  & = & {3 \over 75}
		\barchi^{A}_{0p} + {38\over75}\chi^{S}_{0p}.
	\\
\end{eqnarray}
The coefficients are listed as $C_{NN}$ in Table 2.
The tensor part can similarly be obtained:
\begin{eqnarray}
		\bra \Psi |\calO | \Psi\ket  & = & -{4 \over 75}
		\bartheta^{A}_{0p} + {21\over75}\theta^{S}_{0p} \\
\end{eqnarray}
by using
\begin{eqnarray}
	\bra p^\uparrow p^\uparrow | \sigma^{+}_{3}\sigma^{+}_{6}
	(1+P_{36}^{s}\mp P_{36}^{f}\mp P_{36}^{sf})
	|p^\downarrow p^\downarrow\ket & = &
	{1\over4}\leftw {9\over81}+{9\over81}\mp{17\over81}
	\mp{17\over81}\rightw = \left\{ \matrix{ -{4\over81}
	\cr {13\over81}}\right.
	  \\
	\bra p^\uparrow p^\uparrow | \sigma^{+}_{1}\sigma^{+}_{4}
	(1+P_{14}^{s}\mp P_{14}^{f}\mp P_{14}^{sf})  P_{36}^{sf}
	|p^\downarrow p^\downarrow\ket & = &
	{1\over4}\leftw \!{5\over162}\!+\!{5\over162}\!\mp\!
	{5\over162}\!\mp\!{5\over162}\!\rightw =
	\left\{\matrix{ 0 \cr \!{5\over162}}\right. .
\end{eqnarray}
One can clearly see that the flavor symmetric part of the operator
 is dominant in this state
both for the spin-orbit term and the tensor term.

The actual value is obtained by substituting $\chi^{A}_{0p}$
 by $\barchi_{0p}$
 in eqs.(\ref{eq:barchi0pOGE}) and (\ref{eq:barchi0pIII})
and $\barchi^{S}_{0p}$ by $-{3\over4}\mu_{\rm OGE}$,
and by substituting $\bartheta^{A}_{0p}$ by $\bartheta_{0p}$
in eqs.(\ref{eq:bartheta0pOGE}) and (\ref{eq:bartheta0pIII})
and $\theta^{S}_{0p}$ by $-{1\over8}\mu_{\rm OGE}$.

\newpage

%
%

\newpage

\noindent Figure 1

Mass spectra for a) the negative parity nonstrange baryons and
b) the positive parity ($N$=2) nonstrange baryons.
The observed mass spectrum is shown by stars.
The columns corresponds (from left to right) to
$H_{c}$,  $H_{c}$ + tensor term of OGE, $H_{c}$ + $LS$
and the tensor term of OGE,
$H_{c}$ + $LS$ of OGE for $0d$-quark pairs only,
$H_{c}$ + $LS$ and the tensor terms of OGE and \III\
with $\piii = 0.4$.
Each number corresponds to the spin, $2J$, of the level.

\bigskip
\bigskip
\bigskip

%
%
\noindent Figure 2

Two-nucleon scattering phase shifts for
the spin-triplet relative-$P$-wave states.
The phase shifts are recompiled to present
the strength of the central ($^{3}P_{C}$),
the spin-orbit ($^{3}P_{LS}$)
and the tensor part ($^{3}P_{T}$). (See text.)
The circles corresponds to the experiments VZ40 \cite{SAID},
the solid lines are for $\piii = 0$,
and the dashed lines are for $\piii = 0.4$.

%
%
\newpage
\begin{table}
\caption{Parameters for the quark model
and for the effective meson exchange potential. (See text)
\label{table1}}
\bigskip
\bigskip

\noindent\begin{tabular}{cccccccc}
		\hline
		\hline
		  $m_u$[MeV] & $\alpha_s$ & $V_0^{(2)}$ & $b$[fm] \\
		\hline
		 313 & 1.657 & $-$483.8 & 0.62\\
		\hline
		\hline
		 $\piii$ & $a_{\rm conf}$[MeV/fm] & $V_{\sigma}$[MeV]
		 & $r_\sigma$[fm]& $r_a$[fm] \\
		\hline
		0 & 43.84 & $-$702.9 & 0.617 & 0.25 \\
	  0.4 & 31.40 & $-$528.7 & 0.579 & 0.25 \\
		\hline
		\hline
	\end{tabular}
\end{table}

%
%
%
%
\begin{table}
\caption{Matrix of the OGE and \III\ spin-orbit
and the tensor terms in units of their contribution to the
ground state $\Delta$-N mass difference.
Their contribution to $N$($\FRAC{5}{2}^-$),
$N$($\FRAC{7}{2}^+$)
and the six quark state
for the spin-triplet two-nucleon system
are listed as $C_{N=1}$,  $C_{N=2}$ and $C_{\rm NN}$.
\label{table2}}
\bigskip
\bigskip

\noindent\begin{tabular}{l|rc|ccc}
\hline
\hline
&\rule{0in}{1mm} \hfill OGE\hfill\rule{0in}{1mm}
& {\III} & $C_{N=1}$ &  $C_{N=2}$ &
$C_{\rm NN}$ \\
\hline
$\bra \barcalO_{LS}^A\ket_{0p}$ & ${1\over 2}$
 & $-{1\over 3}{1\over m^2b^2}$
& ${3\over 2}$ & ${3\over 2}$ &  0.12 \\
$ \bra\calO_{LS}^S\ket_{0p}$ & $-{1\over 4}$ & 0
&  & & 1.52 \\
$\bra \barcalO_{LS}^S\ket_{0d}$ & ${1\over 5}$ & 0
& 0 & ${3\over 2}$ &   \\
$ \bra\calO_{LS}^A\ket_{0d}$ & $-{1\over 10}$ & 0
& & &   \\
$\bra \barcalO_{tens}^A\ket_{0p}$ & ${1\over 12}$
& $-{5\over 18}{1\over m^2b^2}$
& $-{3\over 5}$ & $-{3\over 5}$ & $-$0.16\\
$ \bra\calO_{tens}^S\ket_{0p}$ & $-{1\over 24}$ & 0
&   &  & 0.84 \\
$\bra \barcalO_{tens}^S\ket_{0d}$ & ${1\over 30}$ & 0
& 0 & $-{3\over 7}$ &  \\
$\bra \barcalO_{tens}^S\ket_{1s-0d}$ & ${1\over 12\sqrt{10}}$ & 0
& 0 & 0 &  \\
$ \bra\calO_{tens}^A\ket_{0d}$ & $-{1\over 60}$ & 0
&   & &   \\
$ \bra\calO_{tens}^A\ket_{1s-0d}$ & $-{1\over 24\sqrt{10}}$ & 0
&  & &  \\
\hline
\hline
\end{tabular}
\end{table}

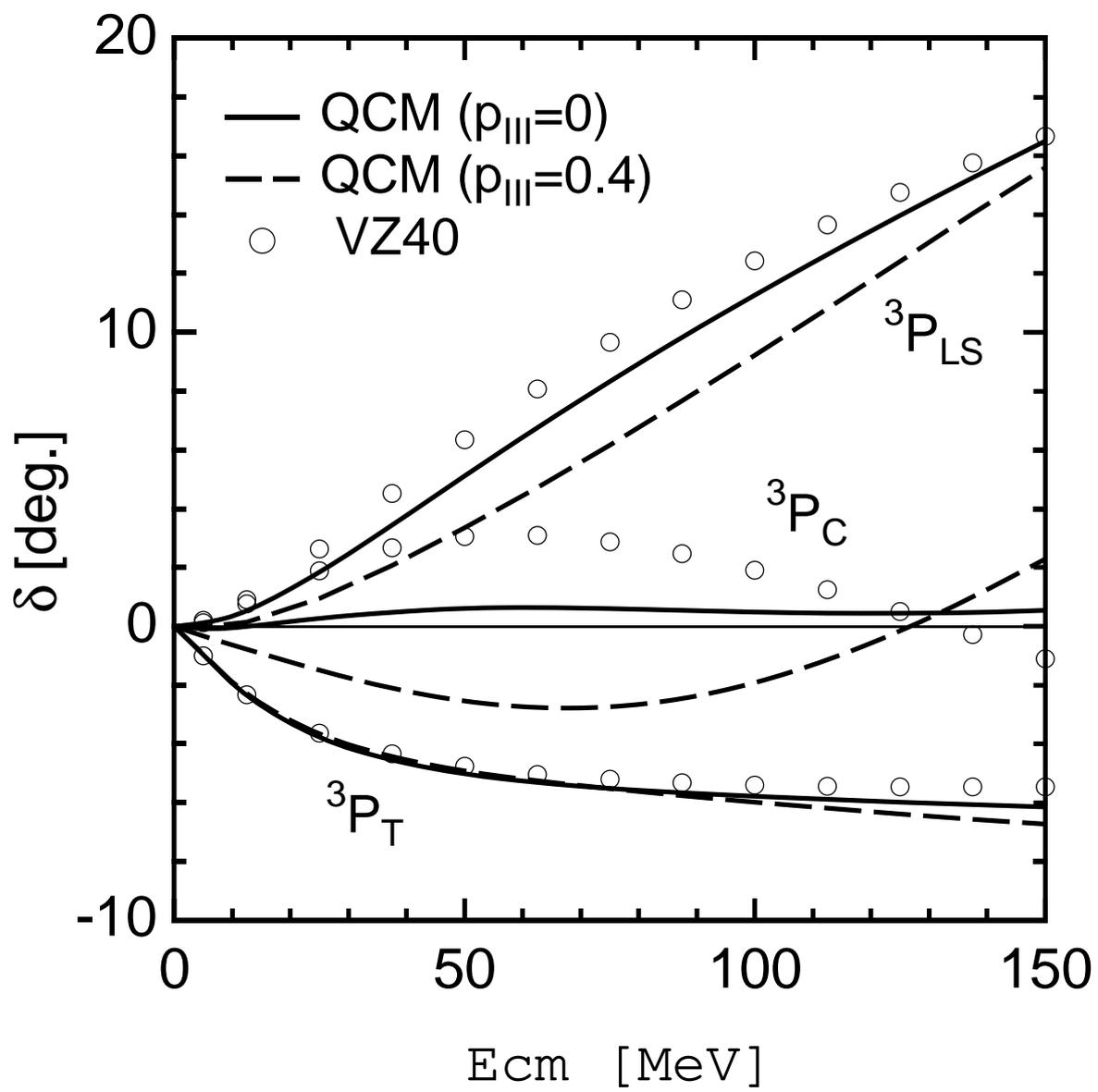
\end{document}